\documentclass[prd,aps,floats,preprint,preprintnumbers]{revtex4}
\usepackage{epsfig}
\newcommand{\be}{\begin{equation}}
\newcommand{\ee}{\end{equation}}
\newcommand{\bea}{\begin{eqnarray}}
\newcommand{\eea}{\end{eqnarray}}
\newcommand{\lag}{{\cal L}}
\newcommand{\vac}{{\rm vac}}
\newcommand{\leff}{\lambda_{\rm eff}}
\newcommand{\mpl}{M_{\rm P}}
\newcommand{\mn}{{\mu\nu}}

\begin{document}
\preprint{EFI-2003-01}
\preprint{SU-GP-03/1-1}
\setlength{\unitlength}{1mm}
 
\title{Can the dark 
energy equation-of-state parameter $w$ \\ be less than $-1$?}

\author{Sean M. Carroll$^{1}$, Mark Hoffman$^{1}$, and Mark Trodden$^{2}$}
\affiliation{$^{1}$Enrico Fermi Institute, Department of Physics, 
and Center for Cosmological Physics,\\
University of Chicago\\
5640 S. Ellis Avenue, Chicago, IL 60637, USA\\
{\tt carroll@theory.uchicago.edu, mb-hoffman@uchicago.edu} \\
\\
$^2$Department of Physics, Syracuse University,\\
Syracuse, NY 13244-1130, USA\\
{\tt trodden@phy.syr.edu}}

\begin{abstract}
Models of dark energy are conveniently characterized by the
equation-of-state parameter $w=p/\rho$, where $\rho$ is the energy
density and $p$ is the pressure.  Imposing the Dominant Energy
Condition, which guarantees stability of the theory, implies that
$w\geq -1$.  Nevertheless, it is conceivable that a well-defined model
could (perhaps temporarily) have $w<-1$ , and indeed such models have
been proposed.  We study the stability of dynamical models exhibiting
$w<-1$ by virtue of a negative kinetic term.  Although naively
unstable, we explore the possibility that these models might be
phenomenologically viable if thought of as effective field theories
valid only up to a certain momentum cutoff.  
Under our most optimistic assumptions, we argue that the instability
timescale can be greater than the age of the universe, but
only if the cutoff
is at or below 100~MeV.  We conclude that it is difficult, although
not necessarily impossible, to construct viable models of dark
energy with $w<-1$; observers should keep an open mind, but
the burden is on theorists to demonstrate
that any proposed new models are not ruled out by rapid vacuum decay.
\end{abstract}
\maketitle


\section{Introduction}
\label{intro}

Cosmological observations strongly indicate that the universe is
dominated by a smoothly distributed, slowly varying dark energy component 
(\cite{Riess:1998cb,Perlmutter:1998np}; for reviews see 
\cite{Carroll:2000fy,Peebles:2002gy}.)  The simplest candidate for
such a source is vacuum energy, or the cosmological constant,
characterized by a pressure equal in magnitude and opposite in
sign to the energy density:
\be
  p_\vac = -\rho_\vac\ .
\ee
While vacuum energy is strictly constant throughout space and time, it is
also worthwhile to consider dynamical candidates for the dark energy.
A convenient parameterization of the recent behavior of any such
candidate comes from generalizing the vacuum-energy equation of state
to
\be
  p = w\rho\ ,
  \label{eqstate}
\ee
which should be thought of as a phenomenological relation reflecting
the current amount of pressure and energy density in the dark
energy.  In particular, the equation-of-state parameter
$w=p/\rho$ is not necessarily constant.  However, given that there
are an uncountable number of conceivable behaviors for the dark energy,
a simple relation such as (\ref{eqstate}) is a useful way to
characterize its current state.  

The equation-of-state parameter is connected directly to the evolution
of the energy density, and thus to the expansion of the universe.
From the conservation-of-energy equation for a component
$\rho_i$ in a Robertson-Walker
cosmology with scale factor $a(t)$ and Hubble parameter
$H=\dot{a}/a$,
\be
  \dot\rho_i = -3H(\rho_i + p_i) \ ,
\ee
it follows that this component evolves with the scale factor as
\be
  {d\ln{\rho_i}\over d\ln{a}} =-3(1+w_i)\ .
\ee
We notice in particular that vacuum energy remains constant, while
the energy density would actually increase as the universe expands
if $w_i < -1$.  The Friedmann equations may be written as
\be
  H^2 = {8\pi G\over 3}\rho -{\kappa \over a^2}\ ,
  \label{f1}
\ee
where $\kappa$ is the spatial curvature, and
\bea
  {\ddot a \over a} &=& -{4\pi G\over 3}(\rho + 3p)\nonumber \\
  &=& -{4\pi G\over 3}(1+3w)\rho \ .
  \label{f2}
\eea
From (\ref{f2}), we see that the universe will accelerate 
($\ddot a > 0$) if $w < -1/3$.  (Of course this is the effective
$w$ of all the energy in the universe; if there is a combination
of matter and dark energy, the dark energy will have to have a 
more negative $w$ in order to cause acceleration.)  From
(\ref{f1}) we see that a flat universe dominated by a component with
constant $w$ will expand as
\be
  a \propto t^{2 / 3(1+w)} \ ,
\ee
unless $w=-1$, for which the expansion will be exponential.
(For $w<-1$, one should choose $t<0$ in this expression.)

What are the possible values $w$ may take?  It is hard to make sweeping
statements about a component of energy about which we know so little.
In general relativity, it is conventional to restrict the possible
energy-momentum tensors by imposing ``energy conditions''.  In
\cite{Garnavich:1998th} it was suggested that a reasonable constraint to 
impose would be the null dominant energy condition, or NDEC
(see Section~\ref{ecs} for discussion).  
The physical motivation for a condition such as the NDEC is to
prevent instability of the vacuum or propagation of energy outside
the light cone.  Applied to an equation of state of the form
(\ref{eqstate}), the NDEC implies $w\geq -1$.  Thus, pure vacuum
energy is a limiting case; any other allowed component would diminish
in energy as the universe expands.

Given our ignorance about the nature of the dark energy, it is worth
asking whether this mysterious substance might actually confound our
expectations for a well-behaved energy source by violating the NDEC.
Given that the dark energy should have positive energy density
(to account for the necessary density to make the universe flat) and
negative pressure (to explain the acceleration observed in the supernova
data), such a violation would imply $w < -1$.
It has been known for some time that such energy components can 
occur~\cite{Nilles:1983ge,Barrow:yc,Pollock:xe}.
Their role as possible dark energy candidates was raised by Caldwell 
\cite{Caldwell:1999ew}, who 
referred to NDEC-violating sources as ``phantom'' components,
and has been since investigated by several authors (for some examples 
see 
\cite{Sahni:1999gb,Parker:1999td,Chiba:1999ka,Boisseau:2000pr,Schulz:2001yx,
Faraoni:2001tq,maor2,Onemli:2002hr,Torres:2002pe,Frampton:2002tu}).  
Observational limits on $w$ \cite{Hannestad:2002ur,Melchiorri:2002ux} are
conventionally expressed as allowed regions in the $w$-$\Omega_{\rm M}$
plane, assuming a flat universe ($\kappa =0$, or $\Omega_{\rm M}
+ \Omega_{X}=1$, where $X$ stands for the dark energy).
Current limits~\cite{Melchiorri:2002ux}, obtained by combining results 
from cosmic microwave background experiments with large scale structure data, 
the Hubble parameter measurement from the Hubble Space Telescope and luminosity measurements of Type Ia supernovae, give $-1.62< w <-0.74$ at the $95 \%$ 
confidence level.

It is straightforward to examine the cosmological consequences of
a dark energy component which is strictly constant throughout
space, and evolving with any value of $w$.  Any physical example of
such a component, however, will necessarily have fluctuations
(so long as $w\neq -1$).  It is therefore important to determine
whether these fluctuations can lead to a catastrophic destablization
of the vacuum.  Unfortunately, because $p=w\rho$ is a phenomenological
description valid for a certain configuration rather than a true
equation of state, specifying $w$ is not enough to sensibly discuss
the evolution of perturbations, since $\delta p\neq w\delta\rho$. 
We must therefore choose a specific model. In particular, the simplest way to
obtain
a phantom component ($w<-1$) is to consider a scalar field $\phi$ 
with {\sl negative} kinetic and gradient energy \cite{Caldwell:1999ew},
\be
  \rho_\phi = -{1\over 2}{\dot\phi}^2 - {1\over 2}(\nabla\phi)^2
  + V(\phi)\ . 
\ee
Fluctuations in this field
have a negative energy, and it may be possible
for the vacuum to decay into a collection of positive-energy 
and negative-energy particles.  If the timescale for such an instability
is less than the age of the universe, the phantom component would not
be a viable candidate for dark energy.

Our goal in this paper is to ask whether phantom components are
necessarily plagued by vacuum instability, and hence whether observers
should take seriously the possibility that $w<-1$.
We will start with a survey of energy conditions and model-independent
considerations in Section~\ref{ecs}.  In Section~\ref{cosmo}, we describe the 
cosmology of $w<-1$ models more thoroughly and
investigate linear perturbations in a cosmological model, demonstrating
that it can be compatible with current observations.  To investigate
stability beyond linear order, in Section~\ref{toy} we consider a
classical toy model of a phantom harmonic oscillator coupled to an
ordinary oscillator, and demonstrate numerically that, for small perturbations 
and sufficiently small values of the coupling constant, there exist both stable and unstable regions of parameter space.   

The heart of our paper is in Section \ref{quantum}, where we
consider the field theory of a phantom component coupled to 
gravity, and calculate the decay rate of a single phantom particle
into several phantoms plus gravitons. 
The rate is naively infinite, due to the infinite phase space of
high-momentum particles.  We argue that, considering the phantom Lagrangian as 
an effective theory, the rate may be rendered finite by
imposing a momentum cutoff and demonstrate that, if we restrict attention to
couplings in the potential, then, for a momentum cutoff not far below the
Planck scale, and for a suitable potential, phantom quanta may be stable
against decay into gravitons and other particles over a timescale long compared
to the age of the universe. However, when we include derivative couplings of
phantom particles to gravitons, we find that such operators can lead to
unacceptably short lifetimes for phantom particles.


\section{Classical energy conditions}
\label{ecs}

In classical general relativity, without having a specific model
for the matter sources, we can nevertheless invoke energy
conditions which restrict the form of the energy-momentum
tensor $T_{\mu\nu}$.  In this section we will briefly review these energy
conditions, discuss which are relevant for cosmology, and compare
them to the condition $w\geq -1$.  A related discussion can be found
in \cite{McInnes:2002qw}.

Each of the energy conditions can be stated in a coordinate-invariant
way, in terms of $T_{\mu\nu}$ and some vector fields of fixed
character (timelike/null/spacelike).  For purposes of physical
insight, it is often helpful to consider the case of a perfect
fluid, for which the energy-momentum tensor takes the form
\be
  T_{\mu\nu} = (\rho + p)U_\mu U_\nu + p g_{\mu\nu}\ ,
\ee
where $\rho$ is the energy density, $p$ the pressure, $U^\mu$ the
fluid four-velocity, and $g_{\mu\nu}$ the metric.  (More precisely,
$\rho$ and $p$ are the energy density and pressure as measured in
the rest frame of the fluid, but the shorthand designations are
standard.)  Our metric signature convention throughout this paper is ($-
$$+$$+$$+$).

\begin{figure}
  \begin{center}
 \includegraphics[scale=0.4]{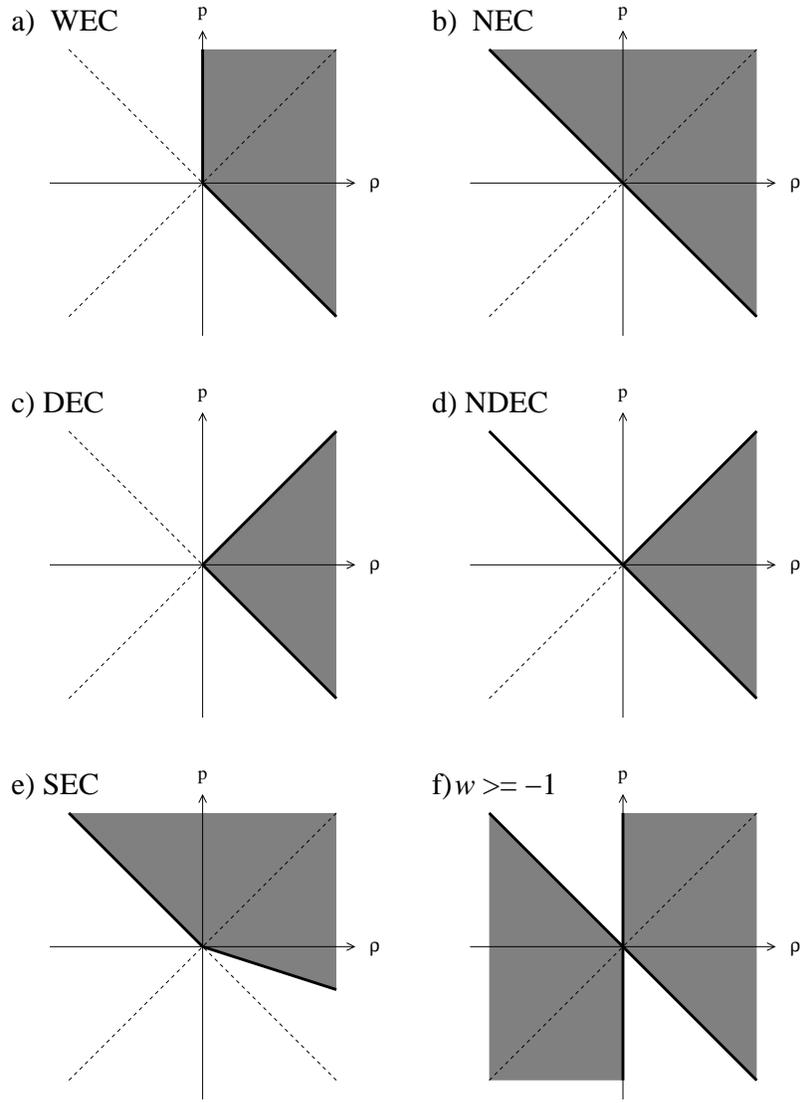}
  \end{center}
  \caption{Shaded regions in the $\rho$-$p$ plane are those which obey the 
  designated energy conditions. Illustrated 
  are the Weak Energy Condition (WEC), Null Energy Condition
  (NEC), Dominant Energy Condition (DEC), Null Dominant
  Energy Condition (NDEC), the Strong Energy Condition (SEC),
  and the condition $w\geq -1$.  Definitions of each condition are
  found in the text.
  }
  \label{ecfig}
\end{figure}

The most common energy conditions are the following:
\begin{itemize}
\item The Weak Energy Condition or WEC
  states that $T_\mn t^\mu t^\nu \geq 0$ for all timelike 
  vectors $t^\mu$, or equivalently that $\rho \geq 0$
  and $\rho + p \geq 0$.
\item The Null Energy Condition or NEC
  states that $T_\mn \ell^\mu \ell^\nu \geq 0$ for all null
  vectors $\ell^\mu$, or equivalently that $\rho + p \geq 0$.
\item The Dominant Energy Condition or DEC includes the WEC
  ($T_\mn t^\mu t^\nu \geq 0$ for all timelike 
  vectors $t^\mu$), as well as the additional requirement that
  $T^\mn t_\mu$ is a non-spacelike vector ({\it i.e.}, that
  $T_\mn T^\nu{}_\lambda t^\mu t^\lambda \leq 0$).  For a perfect
  fluid, these conditions together are equivalent to the 
  simple requirement that $\rho \geq |p|$.
\item The Null Dominant Energy Condition or NDEC is the
  DEC for null vectors only:  for any null vector
  $\ell^\mu$, $T_\mn \ell^\mu \ell^\nu \geq 0$ and
  $T^\mn \ell_\mu$ is a non-spacelike vector.  The allowed density
  and pressure are the same as for the DEC, except that negative
  densities are allowed so long as $p=-\rho$.
\item The Strong Energy Condition or SEC states that
  $T_\mn t^\mu t^\nu \geq {1\over 2}T^\lambda{}_\lambda
  t^\sigma t_\sigma$ for all timelike vectors $t^\mu$, or
  equivalently that $\rho + p \geq 0$ and $\rho + 3p \geq 0$.
\end{itemize}
In Figure \ref{ecfig} we have plotted these conditions
as restrictions on allowed regions of the 
$\rho$-$p$ plane.  We have also 
plotted the condition $w\geq -1$ for comparison.  Note that
$w\geq -1$ is not equivalent to any of the energy conditions, although
it is implied by the WEC, the DEC, and the NDEC.

The different energy conditions are used in different contexts
--- for example, the WEC and SEC are used in singularity theorems --- and
we will not review them in detail here (see \cite{hawkingellis} for
a discussion).  Our present concern is to understand under what
conditions a hypothetical
dark energy component would be guaranteed to be 
stable.  For this purpose, the relevant result is
the ``conservation theorem'' of Hawking and Ellis 
\cite{hawkingellis,Carter:2002wz}.
The conservation theorem invokes the DEC, and uses it to show that
energy cannot propagate outside the light cone; in particular, 
if $T_\mn$ vanishes on some closed region of a spacelike hypersurface, it
will vanish everywhere in the future Cauchy development of that region
--- energy-momentum cannot spontaneously appear from nothing.
A source obeying the DEC is therefore guaranteed to be stable.

For cosmological purposes, however, the DEC is somewhat too restrictive,
as it excludes a negative cosmological constant (which is physically
perfectly reasonable, even if it is not indicated by the data).
This is why \cite{Garnavich:1998th} advocated use of the NDEC in 
cosmology, since the NDEC is equivalent to the DEC except that 
negative values of $\rho$ are allowed so long as $p=-\rho$.  
Of course, the less restrictive NDEC 
invalidates the conservation theorem, as a
simple example shows.  Consider a theory with a negative vacuum
energy $\rho_{\rm vac}<0$, and an ordinary scalar field $\psi$
(not a phantom) with potential $V(\psi) = {1\over 2}m^2\psi^2$.
Then both the vacuum energy and the scalar field obey the NDEC (although
their sum may not, since the NDEC is a nonlinear constraint.)
Imagine a field configuration with $\psi = \sqrt{-2\rho_{\rm vac}/m^2}$
and $\dot\psi = 0$
everywhere throughout space.  The energy-momentum tensor of this configuration
is exactly zero [since $V(\psi)=-\rho_{\rm vac}$], but it would
instantly begin evolving as $\psi$ rolled toward the minimum of its
potential.  There is nothing unphysical about such a situation, 
however, since a stable state is achieved once the field reaches
$\psi=0$.  The NDEC, therefore, does not guarantee stability with the
confidence that the DEC does.  However, it seems legitimate to 
ask that dynamical fields obey the DEC, while the cosmological constant
is allowed as an exception.

The DEC implies that $w\geq -1$.  (Indeed, for cosmological purposes
we are interested in a source with $\rho>0$; in that case,
{\sl all} of the energy conditions imply $w\geq -1$.)
Therefore, if we allow for
pressures which are less than $-\rho$, we cannot guarantee the stability
of the vacuum.  The converse, however, is not true; a phantom component
will not necessarily allow vacuum decay.  In fact, since $p=w\rho$ is just a 
convenient
parameterization and not strictly-speaking an equation of state, the 
issue of stability is
somewhat complicated, as we shall see in the next few 
sections. Nevertheless, it seems 
very likely that energy sources which violate the DEC generally
will be unstable, and this is what we will find in the specific
example considered in Section \ref{quantum}.  Our philosophy, 
therefore, is neither to dismiss the possibility of DEC violation
on the grounds that we cannot prove stability, nor to blithely accept
DEC violation on the grounds that we cannot prove instability, but
instead to see whether it is plausible that the timescale for
instability could be sufficiently long so as to be irrelevant for
practical purposes.
 

\section{Cosmological Evolution and Perturbations}
\label{cosmo}

Consider a flat Robertson-Walker universe with metric
\be
  ds^2 = -dt^2 + a^2(t)[dx^2 + dy^2 +dz^2] \ ,
\ee
for which the Einstein equations are the Friedmann equations
(\ref{f1}) and (\ref{f2}).  The cosmology and fate of a universe
containing an energy component with constant $w<-1$ are
relatively simple and have been examined in, for example,
\cite{Caldwell:1999ew}. As an example, consider the case in which the
universe contains only dust and phantom matter. Then, if the universe
ceases to be matter-dominated at cosmological time $t_m$, then the
solution for the scale factor is
\begin{equation}
\label{constwscalefactor}
a(t)=a(t_m)\left[-w+(1+w)\left(\frac{t}{t_m}\right)\right]^{2/3(1+w)} \ .
\end{equation}
From this 
expression it is easy to see that phantom matter eventually comes to dominate 
the universe and that, since the Ricci scalar is given 
by
\begin{equation}
\label{wconstricciscalar}
R=\frac{4(1-3w)}{3(1+w)^2}\left[t-
        \left(\frac{w}{1+w}\right)t_m\right]^{-2} \ ,
\end{equation}
there is a future curvature singularity at $t=wt_m/(1+w)$. This occurs
because, even though the energy densities in ordinary types of matter
are redshifting away, the energy density in phantom matter increases
in an expanding universe. Thus, the fate of the universe in these
models \cite{Starobinsky:1999yw} may be very different from that
expected
\cite{Starkman:1999pg,Krauss:1999hj,Avelino:2000ix,Gudmundsson:2001gd,
Huterer:2002wf} in $w>-1$ dark energy models.  

It is, however, simple to construct models of phantom energy in which a
future singularity is avoided. As we shall see in this section,
scalar field models can yield a period of time in which the expansion
proceeds with $w<-1$ and yet settles back to $w\geq -1$ (in our case
$w=-1$) at even later times, thus sidestepping the predictions of
$w=$~constant models.
Consider a scalar field theory with action
$S=\int \lag \sqrt{|g|}\, d^4x$, and Lagrange density given
by
\be
  \lag = {1\over 2}g^{\mu\nu}(\partial_\mu\phi)(\partial_\nu\phi)
  -V(\phi)\ .
\ee
The notable feature of this model is that the sign of the kinetic
term is reversed from its conventional value [in our conventions the
usual expression would be $-{1\over2}(\partial\phi)^2$].
The equation of motion for $\phi$ becomes
\be
  \label{phieom}
  \ddot\phi + 3H\dot\phi - a^{-2}\nabla^2\phi - V'(\phi) = 0\ , 
\ee
where $\nabla^2$ is the spatial Laplacian, $\nabla^2\phi =
\partial_x^2\phi + \partial_y^2\phi +\partial_x^2\phi$, and a prime
denotes differentiation with respect to $\phi$.  Here the energy
density $\rho_{\phi}$ and the pressure $p_{\phi}$ for a homogeneous
$\phi$ field are given by
\begin{equation}
\label{rhophi}
\rho_{\phi}=-{1\over 2}\dot{\phi}^2 +V(\phi) \ ,
\end{equation}
\begin{equation}
\label{pphi}
p_{\phi}=-{1\over 2}\dot{\phi}^2 -V(\phi) \ ,
\end{equation}
so that the equation-of-state parameter
\be
  w={p \over \rho} =
  \frac{{1\over 2}\dot{\phi}^2 +V(\phi)}
  {{1\over 2}\dot{\phi}^2-V(\phi)}
  \label{w}
\ee
satisfies $w\leq -1$.

We are interested in the cosmological evolution of this model and in the 
behavior of linearized perturbations of the phantom scalar field in the 
resulting cosmological background.  As noted in \cite{Caldwell:1999ew}, the 
spectrum of fluctuations of a phantom field evolves similarly to that for a 
quintessence field.  Consider metric perturbations in the synchronous
gauge,
\be
  ds^2 = -dt^2 + a^2(\delta_{ij}+h_{ij})dx^idx^j\ .
\ee
A Fourier mode of the phantom field 
\be
  \phi(t,{\bf k}) = {1\over \sqrt{2\pi}} \int \phi(t, {\bf x})
  e^{-i{\bf k}\cdot{\bf x}}\, d^3x
\ee
satisfies the equation of motion
\be
\ddot{\delta\phi_k} + 3H\dot{\delta\phi_k} + (k^2 - V^{\prime\prime})
        {\delta\phi_k}= -\frac{1}{2}\dot{h}{\dot\phi},
\ee
where $h$ is the trace of the synchronous gauge metric 
perturbation $h_{ij}$.  The effective mass for 
the perturbation is $(k^2 - V^{\prime\prime})^{1/2}$.  
On large scales one may 
worry that this effective mass could become imaginary for a positive 
$V^{\prime\prime}$.  We note, however, that a similar problem exists for a 
canonical scalar field for negative $V^{\prime\prime}$, and is thus easily 
avoided by the choice of potential; in particular, we may choose
a potential with $V'' < 0$.  
An analysis of 
perturbations in more general non-canonical scalar field models is given in 
\cite{Garriga:1999}.  In this paper, we will examine
by explicit calculation the evolution of fluctuations in a 
specific model. 

\begin{figure}
  \begin{center}
 \includegraphics[scale=1]{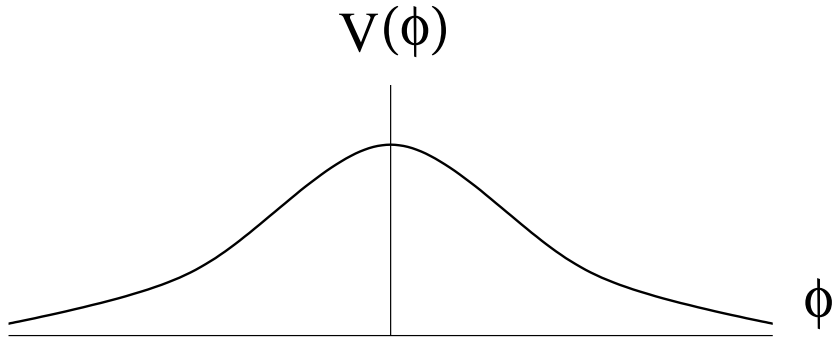}
  \end{center}
  \caption{The gaussian potential energy of (\ref{potential}).
  The phantom scalar will evolve to the top of the hill and 
  oscillate around the maximum.
  }
  \label{gaussian}
\end{figure}
\begin{figure}
  \begin{center}
   \includegraphics[scale=0.4]{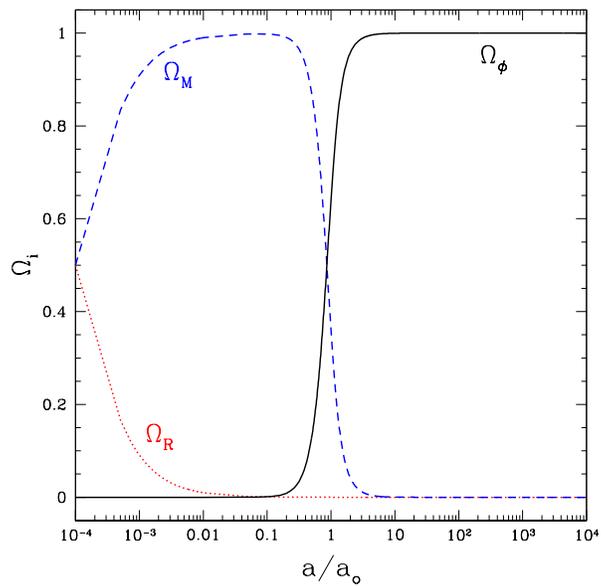}
  \end{center}
  \caption{Evolution of the density parameters in radiation
  ($\Omega_{\rm R}$), matter ($\Omega_{\rm M}$), and the
  phantom field ($\Omega_\phi$).}
  \label{omega_fig}
\end{figure}
\begin{figure}
  \begin{center}
  \includegraphics[scale=0.4]{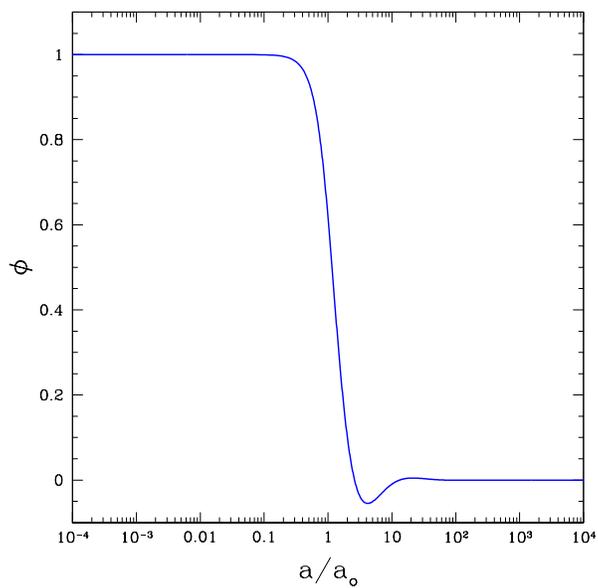}
  \end{center}
  \caption{Evolution of the phantom field $\phi$ as a function of
  the scale factor.}
  \label{phi_fig}
\end{figure}
\begin{figure}
  \begin{center}
  \includegraphics[scale=0.4]{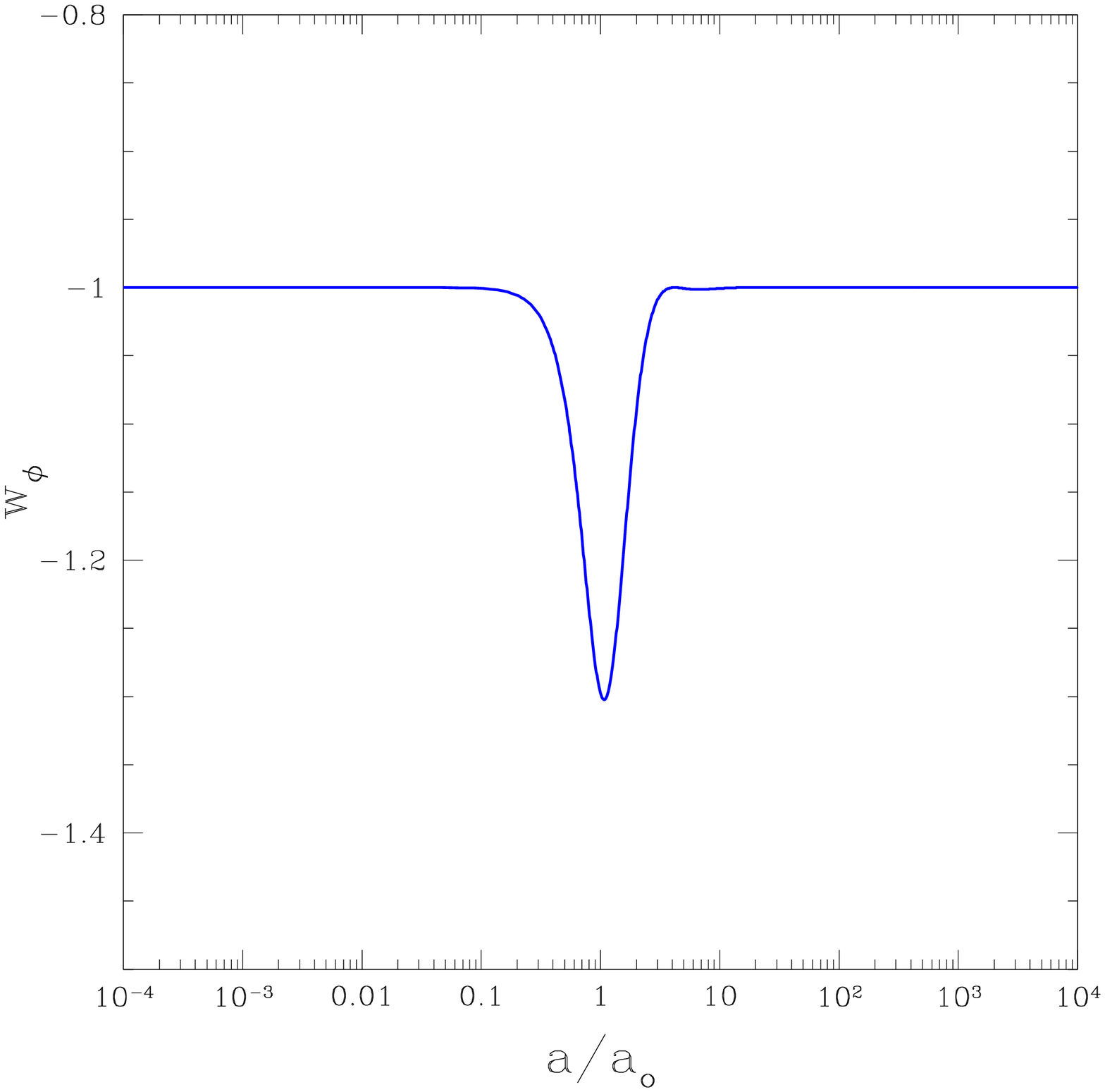}
  \end{center}
  \caption{Evolution of 
the equation of state parameter $w$ for the phantom field as a function of the 
scale factor.
  }
  \label{w_fig}
\end{figure}

Since scalar fields with negative kinetic terms evolve to the {\it maxima} of 
their classical potential, we consider a gaussian potential,
\be
\label{potential}
  V(\phi) = V_0 e^{-(\phi^2/\sigma^2)}\ ,
\ee
where $V_0$ is the overall scale and $\sigma$ is a constant 
describing the width of the gaussian. The potential is represented in
Figure~\ref{gaussian}.  We obtained the cosmological evolution by 
numerically solving the equations of motion~(\ref{f1}), (\ref{f2}), and 
(\ref{phieom}).  The initial conditions were $\phi_{\rm initial}=\mpl$ and
$\dot\phi_{\rm initial}=0$, where $\mpl = (8\pi G)^{-1/2}$ is the reduced
Planck mass.  Since we require dark energy domination at the 
present epoch in the universe, with $\Omega_X \sim 0.7$ we choose $\sigma = 
\mpl$, $V_0 = 3 \mpl^2 H_0^2$. The results are plotted in 
figures~\ref{omega_fig},\ref{phi_fig} and \ref{w_fig}.

Note that during the initial stages of evolution the field $\phi$ is
frozen by the expansion and acts as a negligibly small vacuum energy
component (with $w\simeq -1$). However, at later times the field
begins to evolve more rapidly towards the maximum of its potential,
the energy density in the phantom field becomes cosmologically
dominant, and during this period the equation of state parameter is
much more negative. Finally, in the very late universe, the field
comes to rest at the maximum of the potential and a period of $w=-1$
acceleration begins. Since $w$ is no longer less than $-1$, this
ensures that there is no future singularity; rather, the universe
eventually settles into a de~Sitter phase.

\begin{figure}
  \begin{center}
 \includegraphics[scale=0.4]{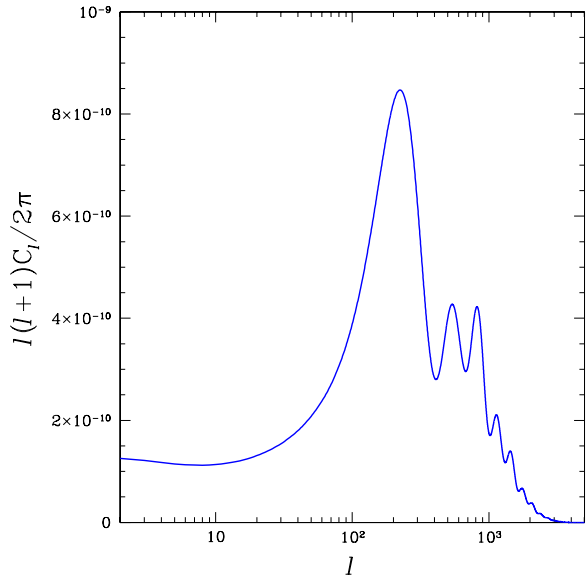}
  \end{center}
  \caption{Predicted multipole moments $C_l$ for cosmic microwave
  background anisotropies.}
  \label{cmb_fig}
\end{figure}

Using the formalism for calculating fluctuations in a general matter field 
developed in \cite{Hu:1998}, one may calculate the effects of the fluctuations 
in this phantom field on the cosmic microwave background radiation (CMB). The 
resulting power spectrum of CMB fluctuations is shown in figure~\ref{cmb_fig}.
We have not attempted a detailed parameter-fitting to CMB data, but
it should be clear that our phantom model with the potential
(\ref{potential}) does not predict any significant
departures from conventional dark-energy scenarios; in particular,
there is no evidence of dramatic instabilities distorting
the power spectrum.  (A more detailed study of the effect on the CMB
power spectrum of phantom fields with a constant $w$ can be found in 
\cite{Schulz:2001yx}.)
However, the formalism for this analysis was based on {\it linear}
perturbation theory, and it is quite plausible that instabilities only
become manifest at higher orders.  In the following sections we 
address this issue, first through numerical investigation of a 
model with two oscillators, and next by calculating the decay rate
of phantom particles into phantoms plus gravitons.
 

\section{Couplings to Normal Matter and Instability}
\label{toy}

Excitations of the scalar field from the model in the previous section
have negative energy.  The existence of negative energy particles may cause
the system to be unstable due to interactions involving these
particles.

In the next section we examine this possibility by estimating the tree
level decay rate of negative energy phantom particles into other
phantoms and gravitons.  Before delving into that calculation, we will
first try to build some intuition about possible instabilities by
considering the classical evolution of a simple system, a coupled pair
of simple harmonic oscillators. Conservation of energy limits the
phase space of a coupled pair of oscillators when both oscillators
have positive energy.  Though the oscillators may exchange energy,
neither oscillator may ever reach an energy greater than the total
initial energy of the system.  Allowing one of the oscillators to have
negative energy removes this limitation on the phase space.  In this
case the positive energy oscillator may increase its energy to any
level so long as the negative energy oscillator decreases its energy
by a compensating amount.  Hence the positive energy oscillator may
reach arbitrarily high energies, and the negative energy oscillator
may reach correspondingly large negative energies.

While conservation of energy does not prevent the oscillators in this
simple model from reaching arbitrarily large energies, there is no
guarantee that the system will be unstable in all regimes.  In fact,
the following analysis will show that for a weak coupling, the
evolution of the energy of each oscillator exhibits a stable
oscillation for an arbitrarily long time.

We consider a coupled pair of 
oscillators, one $\psi$ with positive energy, representing normal matter, and 
the other $\phi$ with negative energy, representing the phantom component.
\be
\label{toy_L}
{\mathcal L} = \left({1\over 2}\dot\psi^2 - m_\psi^2\psi^2\right) 
        - \left({1\over 2}\dot\phi^2 - m_\phi^2\phi^2\right) 
        - \lambda\phi^2\psi^2 \ ,
\ee
where $\lambda$ is a dimensionless coupling constant.
We are interested in 
the evolution of the energy of each oscillator,
\bea
\rho_\psi &=& {1\over 2}\dot\psi^2 + m_\psi^2 \nonumber \\
\rho_\phi &=& - \left({1\over 2}\dot\phi^2 + m_\phi^2\right) \ .
\eea
The equations of motion are
\bea
\label{toy_EOM}
\bar\psi^{\prime\prime} &=& -\left[\left(\frac{m_\psi}{m_\phi}\right)^2
                        + \bar\lambda\bar\phi^2\right]\bar\psi \nonumber \\
\bar\phi^{\prime\prime} &=& -\left[1-\bar\lambda\bar\psi^2\right]\bar\phi \ ,
\eea
where we have defined dimensionless rescaled variables
\bea
  \bar\phi &=& \phi/M\cr
  \bar\psi &=& \psi/M\cr
  \bar\lambda &=& \lambda(M/m_\phi)^2\cr
  \tau &=& m_\phi t\ .
\eea
In this section a prime
denotes diferentiation with respect to the dimensionless time
parameter $\tau$, and $M$ is the scale of the initial
displacements of the oscillators.

\begin{figure}
  \begin{center}
 \includegraphics[scale=0.7]{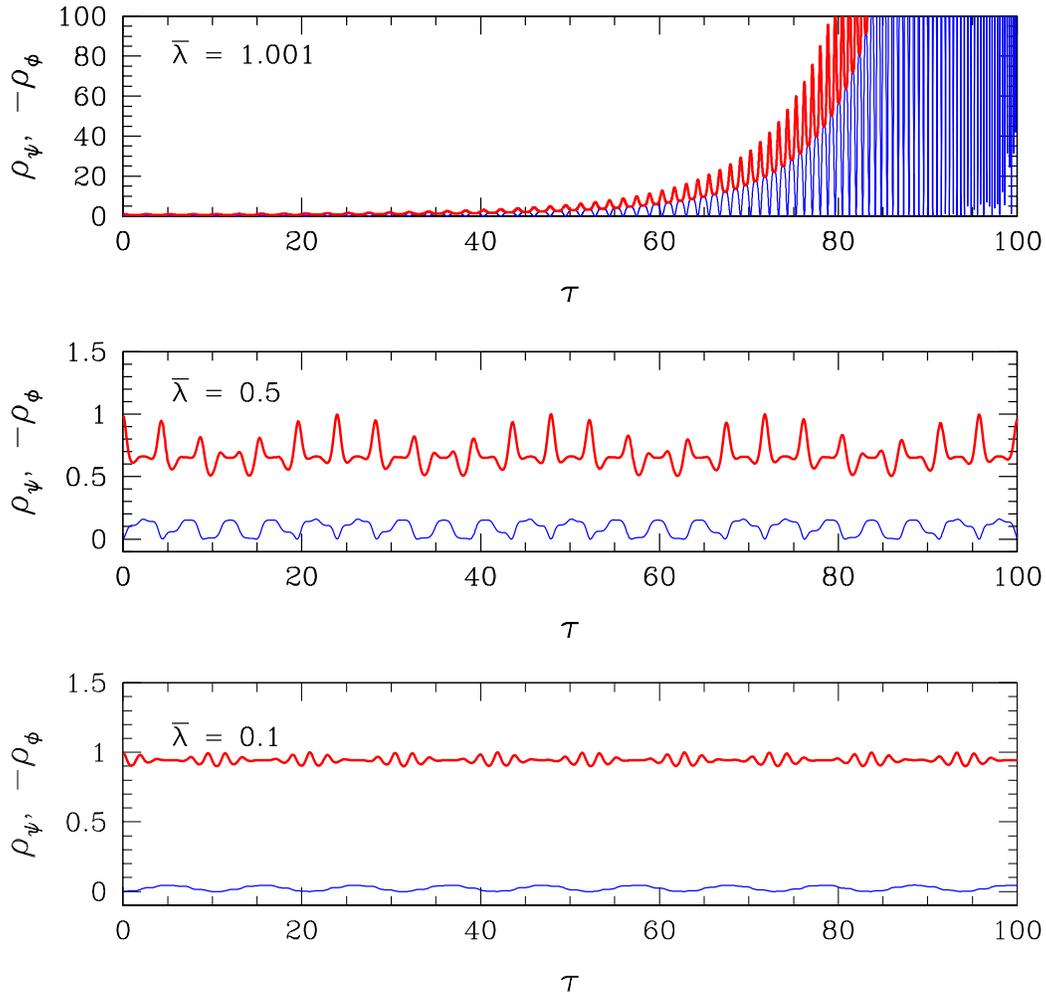}
  \end{center}
\vskip-1cm
\caption{Evolution of the energies of a coupled pair of oscillators, one 
with positive energy, $\rho_\psi$ (thin line), and one with negative energy, 
$\rho_\phi$ (thick line), 
as a function of the rescaled coupling $\bar\lambda$. 
The oscillators are both started at rest displaced a distance $M$
from the origin, and the energies are expressed in units of $(m_\phi M)^2$.  
Note that for 
$\bar\lambda < 1$, $\rho_\psi$ and $-\rho_\phi$ exhibit a stable oscillatory 
behavior, 
while for $\bar\lambda>1$, $\rho_\psi$ and $-\rho_\phi$ rapidly grow.}
  \label{stability_fig}
\end{figure}

We explore the stability of this model by integrating the equations of
motion (\ref{toy_EOM}) numerically for a range of parameters and
initial conditions.  Of particular interest are the simulations with
$m_\psi = 0$, in which one may think of the $\psi$ oscillator as
analogous to the massless graviton.  Three such simulations are shown
in Fig. \ref{stability_fig}.  In each of these integrations the
oscillators are started at rest, displaced a distance $M$ from the
origin.  These plots are similar to those obtained for other initial
conditions, and allowing the $\psi$ field to have a mass does not
qualitatively change the plots shown here.  Note that for $\bar\lambda
< 1$, $\rho_\psi$ and $\rho_\phi$ exhibit a stable oscillatory
behavior, while for $\bar\lambda>1$, $\rho_\psi$ and $-\rho_\phi$
rapidly grow.  From this analysis we conclude that this simple model
is stable for small enough coupling but
exhibits an instability for a large coupling.  While the analysis of
this simple model does not prove that a phantom field in the
cosmological context will be stable to excitations and interactions
with the gravitational field, it does provide some evidence that
stability is possible.

To make some connection to the cosmological model of the previous 
section, we 
consider how small the coupling in this simple model should be in order to have a stable 
solution.  If we imagine a period of early universe inflation, then one expects perturbations in the phantom and gravitational fields to be of order $M \sim 
10^{-5}\mpl$. The mass 
squared of the phantom field is 
\be
m_\phi^2 = \frac{d^2V}{d\phi^2} \sim 
        \frac{V_0}{\mpl^2} \sim (10^{-33} {\rm eV})^2 \ .
\ee
The restriction 
$\bar\lambda < 1$ then implies that the coupling 
in the Lagrangian satisfies $\lambda < 10^{-110}$.  While this at first 
seems to be an 
absurdly small number, we will see in the next section that, given the 
cosmological constraints of the previous section, such a coupling naturally 
results from 
considering perturbations of the phantom and gravitational fields.


\section{Phantom Decay Rate in Field Theory}
\label{quantum}

The toy model of the previous section demonstrated that coupling a
phantom oscillator to an ordinary oscillator results in a system which
may or may not be unstable, depending on the magnitude of the coupling
between them.  It is far from clear, however, that this conclusion
extends immediately to field theory.  Roughly speaking, oscillators
with some frequency correspond to field-theory modes of fixed
wavelength; even if the model is stable when only certain wavelengths
are considered, it does not follow that stability continues to
obtain when integrating over all momenta.

One way of stating this concern is to ask about the decay rate of
single particles that would conventionally be stable.  Because 
excitations of the phantom field have negative energy, we could
imagine a single particle decaying into a large number of phantoms
and ordinary particles.  The rate for this process can be calculated
(at tree level) using ordinary Feynman diagrams.  We will find that
the rate is infinite when arbitrarily high momenta are included, but
can be rendered finite if a cutoff (which might arise, for example,
from higher-derivative terms in the action) is introduced.

Since we are interested in the allowed decay modes of phantom
particles $\phi$ and their associated decay rates, it is instructive to 
first analyze the kinematics of reactions involving phantom fields.
Let us 
adopt the convention that ordinary particles have 4-momenta with positive 
timelike component. Then, since our field has a negative kinetic term, the 
corresponding 4-momentum will have a negative timelike component. This leads to the following useful dictionary for translating between kinematically allowed 
reactions for phantom particles and those for ordinary particles.

Denote 
ordinary particles as $\psi$.  To ask whether a certain reaction is 
kinematically allowed, we can just switch the phantom particles from the right 
side to the left and vice-versa, and ask whether the resulting reaction would 
normally be allowed.  Consider for example the decay of an ordinary particle 
into another particle plus a phantom:
\begin{equation}
 \psi_1 \longrightarrow \psi_2 + \phi \ .
\end{equation}
This will be allowed if the reaction
\begin{equation}
\psi_1 + \phi \longrightarrow \psi_2
\end{equation}
would be allowed by conventional kinematics --- in particular, if the
mass of ordinary particle 2 were greater than the sum of the masses
of ordinary particle 1 and the phantom.  
So it is clear that ordinary particles can
decay into {\it heavier} particles plus phantoms.  For example, if the
electron were coupled to the phantom field, processes such as
\be
  e^- \rightarrow \mu^- + \nu_e + \bar{\nu}_\mu + \phi
\ee
would be allowed.  The muon would then decay back into an electron,
as part of a potentially disasterous cascade.

Next we turn to the decay of phantoms.  
First consider decays into ordinary particles
\begin{equation}
\phi \longrightarrow \psi_1 + \psi_2
\end{equation}
will be allowed if
\begin{equation}
\left|0\right\rangle \longrightarrow \phi + \psi_1 + \psi_2
\end{equation}
would be conventionally allowed, which it is not.  However, consider
a decay into one phantom and one ordinary particle:
\begin{equation}
\phi_1 \longrightarrow \phi_2 + \psi
\end{equation}
will be 
allowed if
\begin{equation}
\phi_2 \longrightarrow \phi_1 + \psi
\end{equation}
would ordinarily be, which requires $m_{\phi_2} > m_{\phi_1} + m_{\psi}$.  So a
phantom can decay into a heavier phantom plus a not-too-heavy ordinary
particle.

If there is only one kind of phantom, one might think it would be stable since 
there would be
no heavier phantoms to decay into. However, several lighter 
particles can mimic the 4-momentum of a heavier particle. Consider the decay of one phantom into two phantoms plus an ordinary particle:
\begin{equation}
\phi_1 \longrightarrow \phi_2 + \phi_3 + \psi \ .
\end{equation}
This will be allowed if
\begin{equation}
\phi_2 + \phi_3 \longrightarrow \phi_1 + \psi
\end{equation}
would ordinarily be, which it aways is for large enough relative
velocities of $\phi_2$ and $\phi_3$.

In summary, if there is only one kind of phantom particle 
(with a unique mass), it can only decay by emitting at least 
two more phantoms, plus at least one ordinary particle.  Ordinary
particles, meanwhile, may decay into phantoms plus other ordinary particles
with a larger effective mass than the original.  A special case
to these rules comes from massless particles.  Although it
would seem kinematically possible for massless particles to decay into
massive particles plus phantoms, massless particles cannot decay in flat space (a reasonable approximation in the backgrounds we are considering)
simply because there is no rest frame in which to calculate the
rate --- no proper time elapses along a null path.

\begin{figure}
  \begin{center}
 \includegraphics[scale=0.8]{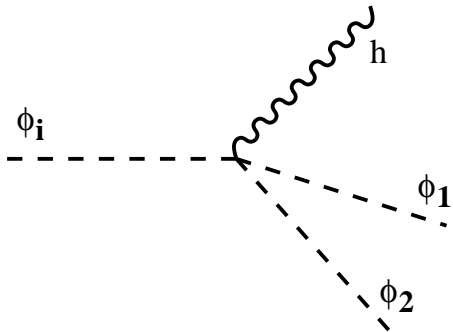}
  \end{center}
  \caption{Decay of a phantom particle into two other 
  phantoms and a graviton}
  \label{diagram}
\end{figure}

Like any other dark-energy scalar field, the phantom should be weakly
coupled to ordinary matter (or it would have been detected through
fifth-force experiments or variation of the constants of nature
\cite{Carroll:1998zi}).  We therefore restrict our attention to only 
gravitons and phantoms.
With the above rules in mind, the phantom decay channel 
involving the smallest number of particles is
\be
  \phi_i \rightarrow h + \phi_1 + \phi_2 \ ,
  \label{phidecay}
\ee
where $h$ is a graviton and $\phi$ is a phantom particle, illustrated in 
figure \ref{diagram}.

Consider the specific model investigated in section~\ref{cosmo},
a phantom scalar with potential
\be
        V(\phi) =
        V_0e^{-\phi^2/\mpl^2}\ .
\ee
We will first consider this potential expanded as a power series 
around some background value $\phi_0\sim \mpl$.
Gravitons, meanwhile, may be represented by
transverse-traceless metric perturbations $g_\mn = g_\mn^{(0)} + h_\mn$,
where $g_\mn^{(0)}$ is the background Robertson-Walker metric.
Strictly speaking, there is no way for a single graviton to couple
non-derivatively to the potential, simply because there is no way
to construct a scalar from a single traceless $h_\mn$.  But to avoid
a surfeit of indices, and because it won't affect the final answer, we will
simply think of the graviton as a dimensionless scalar field $h$; to get
a canonically-normalized field we multiply by $\mpl$.

To study the decay (\ref{phidecay}), we require the interaction
part of the Lagrangian to first order in $h$ and third order in
$\phi$, which is 
\bea
  {\mathcal L}_I &=& \frac{1}{\mpl}(\mpl h)
        \frac{1}{3!}V^{\prime\prime\prime}(\phi_0)\phi^3 \cr\cr
        &=&\leff (\mpl h)\phi^3\ ,
\eea
where
\be
  \leff \equiv 2\phi_0 {V(\phi_0)\over\mpl^5} \sim {V_0\over \mpl^4}
  \sim 10^{-120}\ . 
\ee
The decay rate of a phantom particle through this channel is
\be
\Gamma = \frac{1}{m_{\phi}}\int\frac{d^3p_{h}}{(2\pi)^32E_{h}}
        \frac{d^3p_{\phi_1}}{(2\pi)^32E_{\phi_1}}
        \frac{d^3p_{\phi_2}}{(2\pi)^32E_{\phi_2}}
        |{\mathcal M}|^2(2\pi)^4
        \delta^{(4)}(p_{\phi_i} - p_{\phi_1}-p_{\phi_2}-p_h) \ ,
\ee
where the
matrix element $|{\mathcal M}|$ is just $\leff$ at tree level.  
To get an upper limit on the reaction rate, and hence a lower
limit on the timescale, we assume approximate isotropy, 
so $d^3p \sim |p|^2dp$.  Because the relevant momenta are very
large (and the masses very small), we can also approximate
$E\sim p$.  Putting all of this together we obtain
\be
  \Gamma \sim \frac{\leff^2}{m_{\phi}}\int|p_{h}|dp_{h}
        \int|p_{\phi_1}|dp_{\phi_1}
        \int|p_{\phi_2}|dp_{\phi_2}\delta^{(4)}(p_{\phi_i} - p_{\phi_1}-
        p_{\phi_2}-p_h) \ .
        \label{decayrate}
\ee
If we take the limits on the integrals to be $\infty$, 
the decay rate is clearly infinite.  Hence, the answer to our
investigation into instability seems very clear:  the theory is
dramatically unstable, as individual particles rapidly decay into
cascades of phantoms and gravitons.

However, our philosophy has been to think of this model as an
effective theory valid at low energies.  The reason why the decay
rate diverges is because the phase space is infinite, since the
phantoms can have arbitrarily large negative energies.  Therefore,
this result relies on taking the calculation at face value up to
infinite momentum transfer.  Instead, we should only trust the
phase-space integrals up to some cutoff where new physics might
enter.  For example, we could imagine a higher-derivative term
of the form
\be
  {\mathcal L} \sim -(\partial\phi)^4\ ,
\ee
which would eventually dominate over the negative-kinetic-energy
term we have already introduced.

We have not investigated closely the properties of phantom models
with higher-derivative terms.  Instead, let us crudely approximate
the effect of a momentum cutoff at a scale $\Lambda$ by only
allowing the phase-space integrals to range up to that cutoff.
Looking at (\ref{decayrate}), we can estimate the truncated decay
rate as
\be
  \label{decayrate0}
  \Gamma \sim \leff^2 \frac{\Lambda^2}{m_{\phi}}\ .
\ee
The timescale for decay is $\tau = \Gamma^{-1}$.  Despite any 
fundamental instability, a model will be phenomenologically viable
if the lifetime is greater than the Hubble time $H_0^{-1} \sim
10^{60}\mpl^{-1}$.  Recalling that $m_\phi \sim 10^{-60}\mpl$
and $\leff \sim 10^{-120}$, the lifetime in units of the Hubble time
is
\be
  H_0\tau \sim 10^{120}\left(\frac{\mpl}{\Lambda}\right)^2\ .
\ee
In other words, the lifetime from this decay channel
exceeds cosmological timescales so long as $\Lambda < 10^{60}\mpl$,
which is certainly not a stringent constraint.

However, the infinite phase space for this one decay is not the
only infinity we have to deal with; a single phantom can decay into
arbitrary numbers of gravitons and phantoms, so we must sum over
all the channels.  For each new final-state particle
$\Gamma$ gets multiplied by a factor of $(\Lambda/\mpl)^2$.  The
$\Lambda^2$ comes from the new momentum integral, and the $\mpl^{-2}$
is introduced into $\leff^2$.  So, if we denote the decay 
rate~(\ref{decayrate0}) by 
$\Gamma_0$, then if there are $n$ additional 
particles in the final state the total decay rate is
given by
\be
\Gamma_n = \left(\frac{\Lambda}{\mpl}\right)^{2n}\Gamma_0 \ ,
\ee
Therefore, the total decay rate is 
\be
\Gamma_{\rm total} = \sum_{n=0}^{\infty} (n+1)\Gamma_n \ ,
\ee
where the factor $(n+1)$ comes from the number of different ways the final
state
can be composed of gravitons and phantoms. This yields
\be
\Gamma_{\rm total} = 
        \Gamma_0\left[1-\left(\frac{\Lambda}{\mpl}\right)^2\right]^{-2} \ .
\ee
Thus, the decay rate remains of order $\Gamma_0$ so long as $\Lambda$
is not larger than $\mpl$.  This still seems like a comfortable result.

There is, however, one more possibility to be accounted for.  Part of
the reason we obtained reasonable decay rates from interactions
originating in the potential $V(\phi)$ was because of the small value
of $V_0\sim (10^{-3}{\rm ~eV})^4$.  This suppression is lost if
we consider couplings of gravitons to derivatives of $\phi$.  Since we
are claiming that our model is simply an effective field theory, we
are obligated to include all possible non-renormalizable interactions,
suppressed by appropriate powers of the cutoff scale.  Consider for
example the operator
\be
  {\mathcal L} = {\beta\over \mpl\Lambda}\phi (\mpl h^\mn)
  \partial_\mu\phi\partial_\nu\phi\ ,
\ee
where $\beta$ is a dimensionless coupling.  
(We will use the same cutoff $\Lambda$ for
our non-renormalizable terms as in our momentum integrals; the
$\mpl$ in the denominator comes from normalization of $h^\mn$.) This term
will also contribute to the interaction shown in Figure~\ref{diagram}.  
Following similar logic as above (including two powers of the momentum
in ${\mathcal M}$ from the derivatives), we get
\be
  \Gamma \sim {\beta^2\Lambda^4 \over m_\phi \mpl^2}\ .
\ee
Using $H_0\sim m_\phi \sim 10^{-60}\mpl$, the lifetime in units of
the Hubble time is now
\be
  H_0 \tau \sim \beta^{-2}\left({\mpl\over 10^{30}\Lambda}\right)^4\ .
\ee
Therefore, if $\beta$ is of order unity, to obtain cosmologically
viable decay rates ($H_0 \tau > 1$) we require the cutoff to be
\be
  \Lambda < 10^{-30}\mpl \sim 10^{-3}{\rm eV}\ .
\ee
This is a much smaller cutoff than was required when we considered
coupling through the potential, since the small prefactor $V_0$ is
not around to help us. 

The above result is alleviated somewhat by imposing an approximate
global symmetry on
the theory, that the
Lagrangian density be invariant under $\phi \rightarrow \phi + {\rm constant}$.
Such a symmetry is quite reasonable, as it is the only known way to 
ensure both a nearly-flat potential and appropriately small couplings
to ordinary matter \cite{Carroll:1998zi}.
In this case, the irrelevant operator of lowest dimension leading to phantom
decay into gravitons is
\begin{equation}
{\cal L}= \frac{\gamma}{\mpl^2\Lambda^4}  \left[(\mpl
h^{\mu\nu})\partial_\mu\phi\partial_\nu\phi \right]^2 \ .
\end{equation}
The decay rate is then
\be
  \Gamma \sim {\gamma^2\Lambda^6 \over m_\phi \mpl^4}\ ,
\ee
with an associated lifetime
\be
  H_0 \tau \sim \gamma^{-2}\left({\mpl\over 10^{20}\Lambda}\right)^6\ ,
\ee
which, if $\gamma \sim 1$, leads to the relaxed requirement
$\Lambda<10^{-20}\mpl \sim 100$~MeV.

Under our most optimistic assumptions (of an approximate global
symmetry), we therefore find that the momentum cutoff characterizing
our effective theory must be less than 100~MeV to guarantee that the
instability timescale is greater than the age of the universe.
We find this value to be uncomfortably low, although not necessarily
impossible; dark-energy models are typically characterized by 
energy scales $\sim 10^{-3}$~eV (for the vacuum energy today) and
$\sim 10^{-33}$~eV (for the mass of the scalar field), so perhaps
the scalar field theory is only valid up to relatively low momenta.
Alternatively, one might imagine searching for some mechanism which
would suppress derivative couplings, leaving only the couplings of
gravitons to the potential, which were consistent with a cutoff as
high as the Planck scale.


\section{Conclusions}
There is no doubt that the discovery of a new component of the energy
density of the universe has profound implications for the relationship
between particle physics and gravity. Whether this component be a pure
cosmological constant (whose magnitude we have no idea how to
understand), a dynamical component
\cite{Sahni:1999gb,Parker:1999td,Chiba:1999ka,Boisseau:2000pr,
Schulz:2001yx,
Faraoni:2001tq,maor2,Onemli:2002hr,Torres:2002pe,Frampton:2002tu,Wetterich:fm,
Ratra:1987rm,Caldwell:1997ii,Armendariz-Picon:1999rj,
Armendariz-Picon:2000dh,Armendariz-Picon:2000ah,Mersini:2001su} (whose
special interactions give rise to the tiny vacuum energy we observe)
or an as yet unimagined source, its nature is an outstanding problem
of fundamental physics. The data thus far, although pointing
conclusively to the existence of this dark energy, do not allow us to
distinguish between competing scenarios.  In particular, much of the
allowed parameter space lies in the region $w<-1$ in which some
cherished notion behind our present theories must be sacrificed.

There are several ways in which one may
achieve $w<-1$, including purely negative kinetic terms, non-minimal
kinetic terms, and scalar-tensor theories. In addition, one might
imagine that effective superexponential expansion of the universe
might be obtained by modifying the Friedmann equation. While this is
at odds with a purely four-dimensional general relativistic
description of the universe, such an effect might be obtained in the
context of brane-world models~\cite{Kehagias:1999aa}. 
However, any such modification of the
Friedmann equation must avoid conflict with the precision predictions
of primordial nucleosynthesis~\cite{Tegmark:2001zc,Carroll:2001bv,
Zahn:2002rr}.

In this paper we have taken 
seriously the possibility that $w$ may be less than $-1$ and have asked the 
question:  ``From a particle-physics and general-relativistic point of view, 
can such theories be made consistent?''
In particular, we have considered a specific 
toy model in which the null dominant energy condition is violated and hence 
the resulting space-time may be unstable. In this model the cosmology is 
well-behaved and the theory may be constructed so that it is stable to 
small, linear perturbations.
When we consider higher order effects, however, the model fails to 
remain stable.
The central result of this paper is a field theory calculation of the decay
rate of phantom particles into
gravitons. This decay rate would be infinite if the phantom
theory was fundamental, valid up to arbitrarily high momenta, and would render
the theory useless as a dark energy candidate. We therefore consider the
phantom theory to be an effective theory valid below a scale 
$\Lambda$.  Interestingly, couplings of gravitons to an
appropriate scalar potential do not lead to decay
of phantoms into gravitons and other phantoms on sub-Hubble
timescales, so long as the cutoff $\Lambda$ is below the Planck scale.
However, in such an effective field theory approach, we are mandated
to include in the Lagrangian operators of all possible dimensions, suppressed
by suitable powers of the cutoff scale. In particular, we must include
couplings of gravitons to derivatives of the phantom field. We find that such
operators, even though they may be of high order, can lead to unacceptably
short lifetimes for phantom particles unless the cutoff scale is less than
$100$ MeV, so new physics must appear in the phantom sector at scales
lower than this.

Our analysis demonstrates that a model with
a well-behaved cosmological evolution and stability to linear perturbations 
may still exhibit instability due to higher order interactions. 
We may therefore ask another crucial question: ``Should observers
seeking to constrain cosmological parameters take seriously the 
possibility that $w<-1$?''  Unfortunately the answer is somewhat
ambiguous.  On the one hand, we know of no easy way to construct a
viable model of this sort; on the other, it is certainly conceivable
that new physics in the dark-energy sector kicks in at low scales to
render a phantom model stable, or that phantom behavior is mimicked
by even more exotic mechanisms (such as modifications of the Friedmann
equation).  Therefore, whether or not observations constrain
$w$ to be greater than or equal to $-1$ is still an interesting 
question, although there is a substantial {\it a priori} bias
against the possibility.
For theorists, our conclusion is more straightforward:  the onus is
squarely on would-be phantom model-builders to show how
any specific proposal manages to avoid rapid vacuum decay.


\section*{Acknowledgments}

We would like to thank Roger Blandford, Mark Bowick, Dan Chung, David
Gross, Don Marolf, Laura Mersini and Emil Martinec for useful
conversations.  We especially thank Wayne Hu for his assistance with
calculating power spectra.  MT would like to thank the Kavli Institute
for Theoretical Physics for kind hospitality and support during the
final stage of this project. The work of SC and MH is supported in part by
U.S. Dept. of Energy contract DE-FG02-90ER-40560, 
National Science Foundation grant PHY-0114422 (CfCP), the
Alfred P. Sloan Foundation, and the David and Lucile Packard
Foundation. The work of MT is supported in part by the NSF under grant
PHY-0094122.



\begin{thebibliography}{999}
\parindent=.6em

\bibitem{Riess:1998cb}
A.~G.~Riess {\it et al.}  [Supernova Search Team Collaboration],
Astron.\ J.\  {\bf 116}, 1009 (1998)
[arXiv:astro-ph/9805201].

\bibitem{Perlmutter:1998np}
S.~Perlmutter {\it et al.}  [Supernova Cosmology Project Collaboration],
Astrophys.\ J.\  {\bf 517}, 565 (1999)
[arXiv:astro-ph/9812133].

\bibitem{Carroll:2000fy}
S.~M.~Carroll,
Living Rev.\ Rel.\  {\bf 4}, 1 (2001)
[arXiv:astro-ph/0004075].

\bibitem{Peebles:2002gy}
P.~J.~Peebles and B.~Ratra,
arXiv:astro-ph/0207347.

\bibitem{Garnavich:1998th}
P.~M.~Garnavich {\it et al.},
Astrophys.\ J.\  {\bf 509}, 74 (1998)
[arXiv:astro-ph/9806396].

\bibitem{Nilles:1983ge}
H.~P.~Nilles,
Phys.\ Rept.\  {\bf 110}, 1 (1984).

\bibitem{Barrow:yc}
J.~D.~Barrow,
Nucl.\ Phys.\ B {\bf 310}, 743 (1988).

\bibitem{Pollock:xe}
M.~D.~Pollock,
Phys.\ Lett.\ B {\bf 215}, 635 (1988).

\bibitem{Caldwell:1999ew}
R.~R.~Caldwell,
arXiv:astro-ph/9908168.

\bibitem{Sahni:1999gb}
V.~Sahni and A.~A.~Starobinsky,
Int.\ J.\ Mod.\ Phys.\ D {\bf 9}, 373 (2000)
[arXiv:astro-ph/9904398].

\bibitem{Parker:1999td}
L.~Parker and A.~Raval,
Phys.\ Rev.\ D {\bf 60}, 063512 (1999)
[arXiv:gr-qc/9905031].

\bibitem{Chiba:1999ka}
T.~Chiba, T.~Okabe and M.~Yamaguchi,
Phys.\ Rev.\ D {\bf 62}, 023511 (2000)
[arXiv:astro-ph/9912463].

\bibitem{Boisseau:2000pr}
B.~Boisseau, G.~Esposito-Farese, D.~Polarski and A.~A.~Starobinsky,
Phys.\ Rev.\ Lett.\  {\bf 85}, 2236 (2000)
[arXiv:gr-qc/0001066].

\bibitem{Schulz:2001yx}
A.~E.~Schulz and M.~J.~White,
Phys.\ Rev.\ D {\bf 64}, 043514 (2001)
[arXiv:astro-ph/0104112].

\bibitem{Faraoni:2001tq}
V.~Faraoni,
Int.\ J.\ Mod.\ Phys.\ D {\bf 11}, 471 (2002)
[arXiv:astro-ph/0110067].

\bibitem{maor2}
I.~Maor, R.~Brustein, 
J.~McMahon and P.~J.~Steinhardt,
Phys.\ Rev.\ D {\bf 65}, 123003 (2002).
[arXiv:astro-ph/0112526].

\bibitem{Onemli:2002hr}
V.~K.~Onemli and R.~P.~Woodard,
Class.\ Quant.\ Grav.\  {\bf 19}, 4607 (2002)
[arXiv:gr-qc/0204065].

\bibitem{Torres:2002pe}
D.~F.~Torres,
Phys.\ Rev.\ D {\bf 66}, 043522 (2002)
[arXiv:astro-ph/0204504].

\bibitem{Frampton:2002tu}
P.~H.~Frampton,
arXiv:astro-ph/0209037.

\bibitem{Hannestad:2002ur}
S.~Hannestad and E.~Mortsell,
Phys.\ Rev.\ D {\bf 66}, 063508 (2002)
[arXiv:astro-ph/0205096].

\bibitem{Melchiorri:2002ux}
A.~Melchiorri, L.~Mersini, C.~J.~Odman and M.~Trodden,
arXiv:astro-ph/0211522.

\bibitem{McInnes:2002qw}
B.~McInnes,
arXiv:astro-ph/0210321.

\bibitem{hawkingellis}
S.W. Hawking and G.F.R. Ellis, {\it The Large Scale Structure
of Space-Time}, (Cambridge, Cambridge University Press:  1973).

\bibitem{Carter:2002wz}
B. Carter,
arXiv:gr-qc/0205010.

\bibitem{Garriga:1999} 
J. Garriga and V. Mukhanov, 
Phys.\ Lett.\ B {\bf 458}, 219 (1999) 
[arXiv:hep-th/9904176].

\bibitem{Hu:1998} 
W. Hu, 
Astrophys. J. {\bf 506}, 485 (1998)
[arXiv:astro-ph/9801234].

\bibitem{Starobinsky:1999yw}
A.~A.~Starobinsky,
Grav.\ Cosmol.\  {\bf 6}, 157 (2000)
[arXiv:astro-ph/9912054].

\bibitem{Starkman:1999pg} G. Starkman, 
M. Trodden and T. Vachaspati,
Phys.\ Rev.\ Lett.\  {\bf 83}, 1510 (1999)
[arXiv:astro-ph/9901405].

\bibitem{Krauss:1999hj}
L. M. Krauss and G. D. Starkman,
Astrophys.\ J.\  {\bf 531}, 22 (2000)
[arXiv:astro-ph/9902189].

\bibitem{Avelino:2000ix}
P. P. Avelino, J. P. de 
Carvalho and C. J. Martins,
Phys.\ Lett.\ B {\bf 501}, 257 (2001)
[arXiv:astro-ph/0002153].

\bibitem{Gudmundsson:2001gd}
E. H. Gudmundsson and 
G. Bjornsson,
Astrophys.\ J.\  
{\bf 565}, 1 (2002)
[arXiv:astro-ph/0105547].

\bibitem{Huterer:2002wf}
D. Huterer, G. D. Starkman and 
M. Trodden,
Phys.\ Rev.\ D {\bf 66}, 043511 (2002)
[arXiv:astro-ph/0202256].

\bibitem{Carroll:1998zi}
S.~M.~Carroll,
Phys.\ Rev.\ Lett.\  {\bf 81}, 3067 (1998)
[arXiv:astro-ph/9806099].

\bibitem{Barcelo:2002bv}
C. Barcelo and M. Visser,
arXiv:gr-qc/0205066.

\bibitem{McInnes:2002gf}
B.~McInnes,
arXiv:hep-th/0212014.

\bibitem{Armendariz-Picon:2002km}
C.~Armendariz-Picon,
Phys.\ Rev.\ D {\bf 65}, 104010 (2002)
[arXiv:gr-qc/0201027].

\bibitem{Wetterich:fm}
C.~Wetterich,
Nucl.\ Phys.\ B {\bf 302}, 668 
(1988).

\bibitem{Ratra:1987rm}
B. Ratra and 
P. J. Peebles,
Phys.\ Rev.\ D {\bf 37}, 3406 (1988).
 
\bibitem{Caldwell:1997ii}
R. R. Caldwell, R. Dave and 
P. J. Steinhardt,
Phys.\ Rev.\ Lett.\  {\bf 80}, 1582 
(1998)
[arXiv:astro-ph/9708069].

\bibitem{Armendariz-Picon:1999rj}
C.~Armendariz-Picon, T.~Damour and V.~Mukhanov,
Phys.\ Lett.\ B {\bf 458}, 209 (1999)
[arXiv:hep-
th/9904075].
 
\bibitem{Armendariz-Picon:2000dh}
C.~Armendariz-Picon, V.~Mukhanov 
and P.~J.~Steinhardt,
Phys.\ Rev.\ Lett.\  
{\bf 85}, 4438 (2000)
[arXiv:astro-ph/0004134].
 
\bibitem{Armendariz-Picon:2000ah}
C.~Armendariz-Picon, V.~Mukhanov and 
P.~J.~Steinhardt,
Phys.\ Rev.\ D {\bf 63}, 103510 
(2001)
[arXiv:astro-ph/0006373].

\bibitem{Mersini:2001su}
L.~Mersini, M.~Bastero-Gil and P.~Kanti,
Phys.\ Rev.\ D {\bf 64}, 043508 (2001)
[arXiv:hep-ph/0101210].

\bibitem{Kehagias:1999aa}
A.~Kehagias,
arXiv:hep-th/9911134.

\bibitem{Tegmark:2001zc}
M.~Tegmark,
Phys.\ Rev.\ D {\bf 66}, 103507 (2002)
[arXiv:astro-ph/0101354].


\bibitem{Carroll:2001bv}
S.~M.~Carroll and M.~Kaplinghat,
Phys.\ Rev.\ D {\bf 65}, 063507 (2002)
[arXiv:astro-ph/0108002].

\bibitem{Zahn:2002rr}
O.~Zahn and M.~Zaldarriaga,
arXiv:astro-ph/0212360.

\end{thebibliography}
\end{document}